\documentclass[sigconf,screen]{acmart}
\usepackage{listings}
\usepackage{hyphenat}
\usepackage{graphicx}
\usepackage{textcomp}
\usepackage{xcolor}
\usepackage{algorithm}
\usepackage{algpseudocode}
\usepackage{lipsum}  
\usepackage{caption}
\usepackage{subcaption}
\usepackage{balance}
\usepackage{url}
\usepackage{hyperref}
\usepackage{comment}
\usepackage{booktabs} 

\AtBeginDocument{%
  \providecommand\BibTeX{{%
    \normalfont B\kern-0.5em{\scshape i\kern-0.25em b}\kern-0.8em\TeX}}}

\acmConference[FSE'24]{32nd ACM Symposium on the Foundations of Software Engineering}{July 15--19, 2024,}{ Porto de Galinhas, Brazil}

\acmBooktitle{Companion Proceedings of the 32nd ACM Symposium on the Foundations of Software Engineering (FSE '24), July 15--19, 2024, Porto de Galinhas, Brazil}

\begin{document}

\title{Towards an Engineering Discipline for Resilient Cyber-Physical Systems }
\author{Ricardo D. Caldas}
\email{ricardo.caldas@chalmers.se}
\affiliation{%
  \institution{Chalmers University of Technology, Gothenburg, Sweden}
  \country{}
}

\begin{abstract}
Resilient cyber-physical systems comprise computing systems able to continuously interact with the physical environment in which they operate, despite runtime errors. The term resilience refers to the ability to cope with unexpected inputs while delivering correct service. Examples of resilient computing systems are Google's PageRank and the Bubblesort algorithm. Engineering for resilient cyber-physical systems requires a paradigm shift, prioritizing adaptability to dynamic environments. Software as a tool for self-manag-\\ement is a key instrument for dealing with uncertainty and embedding resilience in these systems. Yet, software engineers encounter the ongoing challenge of ensuring resilience despite environmental dynamic change. My thesis aims to pioneer an engineering discipline for resilient cyber-physical systems. Over four years, we conducted studies, built methods and tools, delivered software packages, and a website offering guidance to practitioners. This paper provides a condensed overview of the problems tackled, our methodology, key contributions, and results highlights. Seeking feedback from the community, this paper serves both as preparation for the thesis defense and as insight into future research prospects.
\end{abstract}    


\maketitle 

\section{Introduction}\label{lb:intro}

Cyber-Physical Systems (CPSs) unite the digital and physical worlds, increasingly supporting individuals and groups in their social and professional endeavors. In contrast to traditional embedded systems, CPSs are often designed as networks of interactive and dynamic elements~\cite{fitzgerald:2014}, including healthcare systems, mobility systems, process control systems, and robotics. Such applications directly reflect strategic economic and social development areas, i.e., transport, energy, well-being industry, and infrastructure. 
The European roadmap and strategy for cyber-physical systems (CyPhERS~\cite{schatz:2015}) lists five challenges to the development of CPSs: interoperability
, autonomy
, privacy
, resilience\footnote{Schatz~\cite{schatz:2015} originally proposes dependability. Instead, we discuss \emph{resilience}.}
, and uncertainty
. Although all these challenges are important and interconnected, we focus on discussing resilience. 

Resilience is {\it ``the persistence of service delivery that can justifiably be trusted, when facing changes.''~\cite{laprie:2008:resilience}}.
Archetypal examples of resilient computing systems are Google's PageRank~\cite{cohen2000resilience,ghoshal2011ranking} and the Bubblesort algorithm when compared to more efficient MergeSort or QuickSort~\cite{ackley2013beyond}.

Engineering resilient CPSs involves rethinking their design focused on dynamic environments. CPSs are subject to interactions with (human) users and operators
~\cite{gil:2020:hil}, changing needs~\cite{bencomo:2010:requirements}, and heterogeneous components joining and leaving the system~\cite{givehchi:2017}. Therefore, the software controlling the CPS operation must automatically {\it manage} dynamic interactions and uncertain environments
~\cite{angelika:2017:decentralized,weyns:2021:agenda},
with {\it guarantees} of compliance with the system requirements~\cite{lemos:2017:assurance:challenges, weyns:2019:perpetual}.
Software engineers, however, face the challenge of guaranteeing that complex systems (e.g., CPS) are dependable to dynamic changes in the system and the environment~\cite{ratasich2019roadmap}.
\looseness=-1

Our study argues that resilience must be embedded in CPS by design and through testing.
Therefore, we aim to develop a software discipline focusing on software-centered resilience design for CPS, utilizing software-aided verification and validating CPS with realistic scenarios. To this end, we developed open-access artifacts that solve fundamental barriers to engineering resilient CPS: scientific studies~\cite{caldas2020hybrid, caldas2021towards, queiroz2024driver, rodrigues2022architecture, rizwan2023ezskiros, caldas2024guidelines, menghi2024diagnostics, silva2024test, joao2024explainability}, scientific artifacts~\cite{gil2021body, askarpour2021robomax}, software packages~\cite{lesunbBSN, rizwanEzSkiROS, rodrigoGeoScenario, AskarpourRoboMAX}, and a website with guidelines~\cite{rosrvftVerificationROSbased}.
Our research aims to significantly contribute to Software Engineering by focusing on resilient CPS as a future software-centered discipline~\cite{klint2005toward,mili1999toward,kazman2000toward,shaw1990prospects}

\begin{figure*}[ht!]
    \centerline{\includegraphics[width=\textwidth]{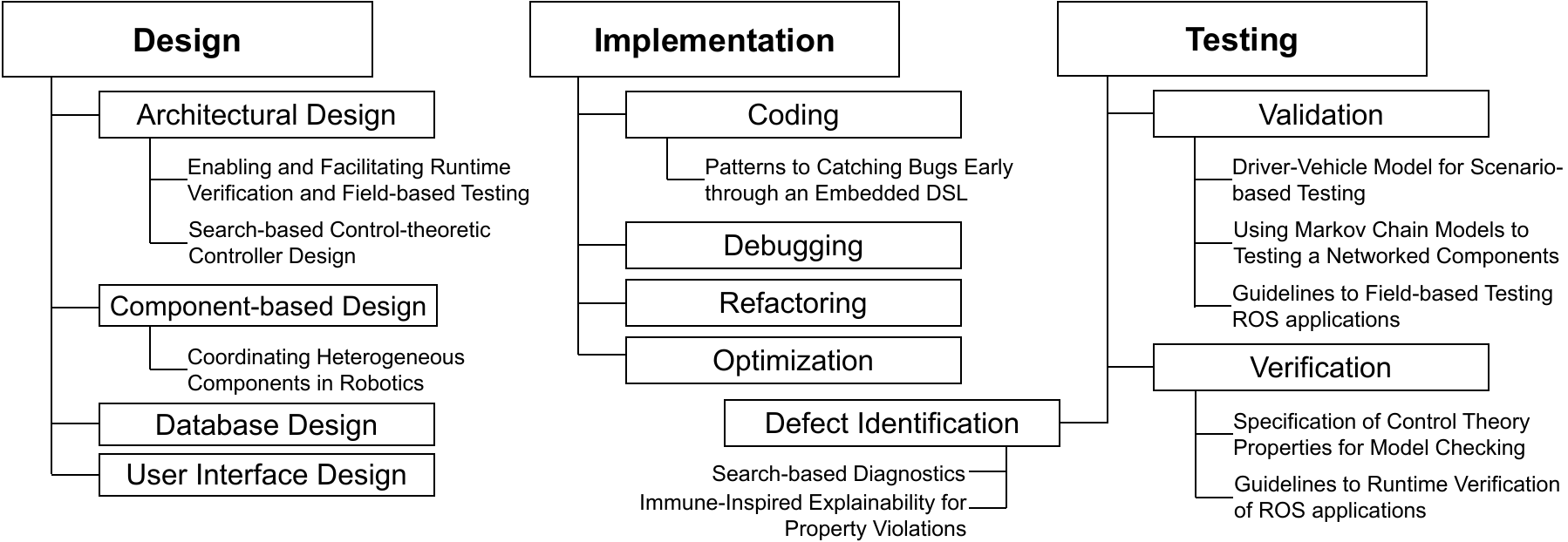}}\vspace{-0.3cm}
    \caption{Overview of our contributions within the activities involved in engineering resilient CPS. \vspace{-0.3cm} }
    \label{fig:contributions}
\end{figure*}

\section{Methodology}\label{lb:methodology}

 Our research goal is to step towards addressing the challenge of ensuring CPS resilience amidst dynamic system and environmental changes. 
To this end, we developed and delivered open-access artifacts that solve fundamental barriers to engineering resilient CPS. We tackled challenges stemming from designing, implementing, and testing resilient CPS.
The artifacts resulted from an incremental and cyclic process 
named {\it design science}
~\cite{hevner:2021}, the process consists of attaining awareness about a problem, devising a solution, and validation. 
To attain {\it awareness} about the problem we used expansive inquiry combined with literature reviews~\cite{kitchenham2009systematic}, and mapping studies~\cite{petersen2008systematic}. Then, we developed {\it solutions} by creatively composing existing scientific artifacts. 
Finally, to {\it validate} 
the creative process employed in the solution step 
we mainly used experimentation~\cite{wohlin2012experimentation} and statistical analyses \cite{de2019evolution, arcuri2014hitchhiker}. 
 Lessons learned and reflections from a validation effort were used to sharpen recurring cycles of awareness attainment and solution refinement.
\looseness=-1

\section{Contributions and Related Work}
We provided the artifacts supporting key activities in establishing a software discipline for resilient CPS engineering depicted in Fig.~\ref{fig:contributions}. This section expands on how our contributions answered research questions to mitigate barriers that hamper research towards the engineering discipline. 

{\bf RQ1: How to design efficient self-adaptation for CPSs?} Designing for fail-safe operation requires enabling {\it self-adaptation} to adequate the system behavior to the specification goals continuously
~\cite{angelika:2017:decentralized,weyns:2021:agenda}. 
To this end, software engineers must design the system to monitor, reason, and act to change itself~\cite{kephart2003vision}. Yet, lack of engineering knowledge, large input spaces, and component coordination threaten such a design. Consequently, the design of efficient self-adaptation requires design decisions at the architecture level and component level. 
At the architecture level, we contributed to enabling and facilitating dynamic analysis processes, i.e., runtime verification and field-based testing. To this end, we prepared guidelines~\cite{caldas2024guidelines,rosrvftVerificationROSbased} that discuss constraint identification
, design patterns
 and instrumentation techniques
. To manage large adaptation space posed by the uncertain environment, we proposed a hybrid architecture for search-based control-theoretic controller design~\cite{caldas2020hybrid}. Differently from other works that use learning, online or offline, and other statistical methods to reduce the adaptation space
~\cite{gerostathopoulos2018adapting,quin:2019,jamshidi2017transfer}
, improves the efficiency of synthesizing adaptation strategies by optimizing the search process exploring the adaptation space. Our approach proved efficiency in three adaptation scenarios on a healthcare system, the BSN~\cite{gil2021body,lesunbBSN}.
At the component level design~\cite{noauthor_robmosys_2017,kaupp_building_2007} ,
we proposed Mission Control~\cite{rodrigues2022architecture} leveraging ensembles to coordinate heterogeneous CPSs components to solve complex missions through efficient coalition formation. Our work evaluated Mission Control using the RoboMAX scenarios~\cite{askarpour2021robomax,AskarpourRoboMAX} to show efficiency in dealing with uncertain scenarios.
\looseness=-1

{\bf RQ2: How to formally ensure whether the adaptive CPS complies with the specifications?} Formally ensuring adaptive behavior requires code-level interventions, formal verification, and defect identification. 
Model-based code-level interventions~\cite{Buch_2014,Coste_Maniere,Kunze_2011} enforce environment assumptions in code but demand an additional modeling step. We introduced four design patterns for embedded domain-specific languages~\cite{wkasowski2023domain} to offer code-level assurances to CPSs~\cite{rizwan2023ezskiros,rizwanEzSkiROS}. 
We also contributed a mapping between software properties in LTL (Linear Temporal Logic) and control quality attributes~\cite{caldas2021towards} as an enabler to checking control-theoretic properties in code. Differently from C\'amara et al.~\cite{camara:2020:bridging}, our mapping technique relies on property specification patterns. We recommend how to generate monitors, prepare the execution environment, instrument the system, execute the system, and post-mortem analysis\cite{caldas2024guidelines,rosrvftVerificationROSbased}. 
Regardless, we proposed a search-based algorithm that takes an execution trace and a violated property to find a diagnostic for what has violated the trace~\cite{menghi2024diagnostics}. Unlike works that extrapolate information from trace-slicing
~\cite{ferrere2015trace,dou2018} 
or showing common behaviors~\cite{luo2014rv,dawes2019}, our diagnostics algorithm explains violations through the mutations applied to the property. In another work, we proposed an immune-inspired negative selection algorithm for detecting candidate features in a violated property~\cite{joao2024explainability}. 

{\bf RQ3: How to validate large and complex CPS operating in realistic scenarios?} CPSs are often large and complex systems and must handle complex human interactions and making test campaigns hard to implement, even harder to automate~\cite{bertolino2021survey}. For this reason, testing CPS in realistic scenarios is costly and slow. The activity requires rigorous planning yet there are no best practices to support field-based testing campaigns~\cite{hochgeschwender2022testing}. 
Consequently, we derived recommendations on how to specify the (un-)desired behavior, generate test cases and prepare oracles, instrument the CPS, collect data, and generate reports\cite{caldas2024guidelines,rosrvftVerificationROSbased}.
Also, we developed human-vehicle models for scenario-basted testing~\cite{queiroz2024driver,rodrigoGeoScenario}. 
In complement to current vehicle simulations, e.g., macroscopic~\cite{sewall:2010:conttrafficsim} or microscopic models~\cite{chao:2020:trafficsimsurvey}, our human-vehicle models compose maneuver models consisting of accurate algebraic implementations. Additionally, we contributed PASTA, a testing technique that uses markov models to simulate patients by considering sensor data trends captured by a networked healthcare system~\cite{silva2024test}. Our method found bugs in the BSN~\cite{gil2021body,lesunbBSN} more efficiently than random generation. 
\looseness=-1

\section{Conclusion} \label{lb:concl}
Our research aims to significantly contribute to Software Engineering by focusing on CPS resilience as a future software-centered discipline. To this end, we followed design science to develop methods and tools to design for efficient self-adaptation, formally ensure that the CPS complies with the specifications, and validate the CPS in realistic scenarios. Ultimately, our study benefits researchers and practitioners in Software Engineering.
\looseness=-1

\bibliographystyle{ACM-Reference-Format}
\bibliography{main}

\end{document}